\documentclass[letter]{spie}  
\usepackage{overcite,graphicx}

\title{Single Photon Source with Individualized Single Photon Certifications}

\author{A. L. Migdall\supit{a}, D. Branning\supit{b}, S.
Castelletto\supit{a,c} and M. Ware\supit{a}
\skiplinehalf
\supit{a}Optical Technology Division,\\
   National Institute of Standards
and Technology,Gaithersburg,
Maryland 20899-8441\\
\supit{b}University of Illinois at Urbana-Champaign,\\
\supit{c}Istituto Elettrotecnico Nazionale G. Ferraris, Turin Italy
}

\authorinfo{Further author information: (Send correspondence to A. L.
Migdall)\\A. L.
Migdall: E-mail: amigdall@nist.gov, Telephone: 1 301 975 2331, FAX: 1
301 869 5700\\ }

\begin{document}
    \maketitle

\begin{abstract}

As currently implemented, single-photon sources cannot be made to
produce single photons with high probability, while simultaneously
suppressing the probability of yielding two or more photons.
Because of this, single photon sources cannot really produce
single photons on demand. We describe a multiplexed system that
allows the probabilities of producing one and more photons to be
adjusted independently, enabling a much better approximation of a
source of single photons on demand. The scheme uses a heralded
photon source based on parametric downconversion, but by
effectively breaking the trigger detector area into multiple
regions, we are able to extract more information about a heralded
photon than is possible with a conventional arrangement. This
scheme allows photons to be produced along with a quantitative
``certification'' that they are single photons. Some of the
single-photon certifications can be significantly better than what
is possible with conventional downconversion sources (using a
unified trigger detector region), as well as being better than
faint laser sources. With such a source of more tightly certified
single photons, it should be possible to improve the maximum
secure bit rate possible over a quantum cryptographic link. We
present an analysis of the relative merits of this method over the
conventional arrangement.

\end{abstract}




\section{INTRODUCTION}
\label{sect:intro}  

The advent of photon-based quantum cryptography, communication and
computation schemes
\cite{BEB84,BEB85,BEB89,BBB91,EKE91,BEN92,BBM92,ERT92,TBZ00,KLM01}
is increasing the need for light sources that produce individual
photons. It is of particular importance that these single photons
be produced in as controlled a manner as possible, as unwanted
additional photons can render quantum cryptographic links insecure
and degrade quantum computation efficiencies \cite{BLM00}. The
ultimate goal is to have completely characterized single photons
produced on demand. More specifically, it would be very useful to
have these photons created in response to an external trigger, or
even to have them arrive repetitively at a fixed, but selectable
rate.

Single photons (or more precisely, approximations thereof) are now
commonly created via the process of parametric downconversion
(PDC) \cite{BUW70,ERT92,SAW99,JSW00,NPW00,GRT01}, although
attenuated lasers and quantum-dots and other single quantum site
sources are also used \cite{JAF96, BHK98, KBK99, BHL00, MIC00,
MIC00b, YKS01}. The PDC process employs a nonlinear optical
crystal that allows individual photons from a pump laser beam to
be converted into pairs of highly correlated photons. Because PDC
creates photons in pairs, the detection of one photon indicates,
or ``heralds'', the existence of its twin, a significant advantage
over other methods (even aside from the potential to directly
produce entangled states). In addition, because the PDC process is
governed by the constraints of phase matching, it is possible to
know the output trajectory, polarization, and wavelength of that
heralded photon.  While PDC has a long history as a ``single
photon source'' and there have been many recent improvements in
efficiency\cite{KWW99}, the scheme has a couple of problems. The
conversion process is random, so while an output photon is
heralded by its twin, there is no control or prior knowledge of
when the heralding event will occur. In addition, there is a
possibility of producing more than one pair at a time and because
that probability increases nonlinearly with the one-photon
probability, one must operate at low one-photon probabilities. So
to be assured that more than one photon is not produced, one must
operate where it is most likely that no photon is produced at all
\cite{NIC00}!

The faint laser scheme suffers the same difficulty as the PDC
method, in possibly producing more than one photon at a time, with
the added disadvantage of not having any herald at all
\cite{BHK98}. Quantum-dot sources offer promise as a new way of
definitively producing single photons, although it remains to be
seen whether their output/collection efficiencies can be made to
approach unity in practice, a requirement for a true on-demand
source.

To surmount the problem of random production in PDC, one uses a
pulsed laser to pump the nonlinear crystal (see for instance
\cite{BHL00}). With a pulsed source, photon pairs can only be
produced at certain times ie., during the laser pulse.
Unfortunately the multiple photon emission problem remains; a high
probability ($P(1)$) of producing a single photon pair during each
pulse leads to an increased likelihood ($P (>1)$) of producing
more than one photon pair during each pulse, defeating the
original goal of having a source of single photons. This problem
occurs regardless of whether sources with Poisson or Bose-Einstein
statistics are used \cite{GRT01,WAM95}. Because of this problem,
pulsed systems also are usually operated with low probability of
producing an output photon pair during a laser pulse (typically
$P(1)\approx$ 0.1 to 0.3) \cite{JAF96, BHK98, BHL00}. Thus, while
photons can only come during specific time windows, most of these
time windows will contain no photons at all, illustrating the
trade off between the two requirements of producing a photon on
demand and being assured that there is, in fact, just one photon.

The improvement presented here allows these two competing
requirements to be adjusted independently by decoupling $P(1)$ and
$P(>1)$. We can then select both the desired likelihood of
production of a photon pair and the desired suppression of
multiple pair events. This is accomplished using an array of
downconverters and detectors (Fig. 1). All of the downconverters
are pumped simultaneously by the same laser pulse. The pump laser
power is chosen so each downconverter has some small probability
of producing a photon pair, while the number of downconverters is
chosen so there is a high likelihood of at least one pair being
created somewhere in the array. The detector associated with each
downconverter allows us to determine which of the downconverters
has fired. This information is used to control an optical
switching circuit directing the other photon of the pair onto the
single output channel. This arrangement allows a much truer
approximation of a single photon on-demand source than is possible
with other methods.

The method imposes a number of requirements on the components used
in a practical implementation. First, a very fast optical switch
is needed, because there must be an optical delay that provides
enough time to throw the appropriate switch before the photon
arrives. Because light travels about 0.3 m/ns in air, a switching
transition time of 10 ns or better would enable a reasonably
compact tabletop setup \cite{PJF02}. Fortunately, there exist
electro-optic switches, such as those based on KDP and LiNbO$_{3}$
that are capable of switching in just a few ns. (The delay time
between an electrical input trigger pulse and the optical
switching transition should also be of this order. This may be a
more difficult, but not impossible, requirement to satisfy.) These
are binary switches (1 input -- 2 outputs) that work by controlled
rotation of the polarization of the light followed by a polarizing
beam splitter. Thus, to set up a circuit to take, for example, 8
input lines and allow one to be selectively switched to the output
line would require a circuit path 3 binary switches deep. Of
course, the laser pulse duration must be shorter than the timing
resolution of the system. In addition to fast switch times, it is
important that the overall optical losses be low, so that there is
a low probability of losing that heralded photon before it can
exit the apparatus; otherwise we would be back to the original
problem of not being sure that we will get a photon. The losses in
a system built of discrete components would include absorptance
and reflectance losses for each component. All of the components
that would be used in this system are existing devices that can be
made with low absorptance losses ($<$ 0.01 \%) and antireflection
(AR) coatings ($\approx$ 0.5 \%). A typical system would consist
of a downconversion crystal (1 AR surface), a delay (0 or 2 AR
surfaces and 0 or 2  mirror surfaces), and an 8-to-1 switch
circuit (4 AR surfaces per 2-to-1 switch), so that the  loss for a
complete system would have less than 15 AR coated surfaces with a
total net loss less than 1-0.995$^{15}$= 7.2 \%, allowing
conservatively a 92.8 \% probability of emitting the heralded
photon. In addition to a single photon on demand system made of
discrete components, it is also of obvious interest to produce a
system made of integrated components. While it is likely that such
a system would not achieve as low a loss as a discrete component
arrangement, a useable system should be possible, although an
engineering effort would be needed to determine the ultimate
limit.

A simple extension of this arrangement can produce a regularly
spaced series of single photons. By ganging up a series of these
``single photon on demand setups'' with additional optical
switches and a series of optical delays, it is possible to produce
an arbitrary length train of single photons. A somewhat more
efficient version of this last scheme could be implemented by
using the array of downconverters and detectors as a single large
pool of photon sources, instead of breaking them up into
subgroups. In this arrangement, detectors that fire indicate which
of the many downconverters have produced photon pairs. This
information is sent to a central control circuit for the optical
switches that then sequences the incoming single photons to
produce the requested output pulse train. By producing a pulse
train long enough to last until the next pump laser pulse arrives,
one can create a continuous train of single-photon pulses
synchronized with an external signal.

These basic concepts could be used to produce higher order photon
number states as well. By using detectors with the capability of
sensing the number of photons in a single pulse, the switching
circuit could just as well direct the outputs of those converters
that produced multiple photon pairs to the output channel. That
would result in an output pulse train with each pulse containing
the desired number of photons.

\begin{figure}
  \centerline{\includegraphics{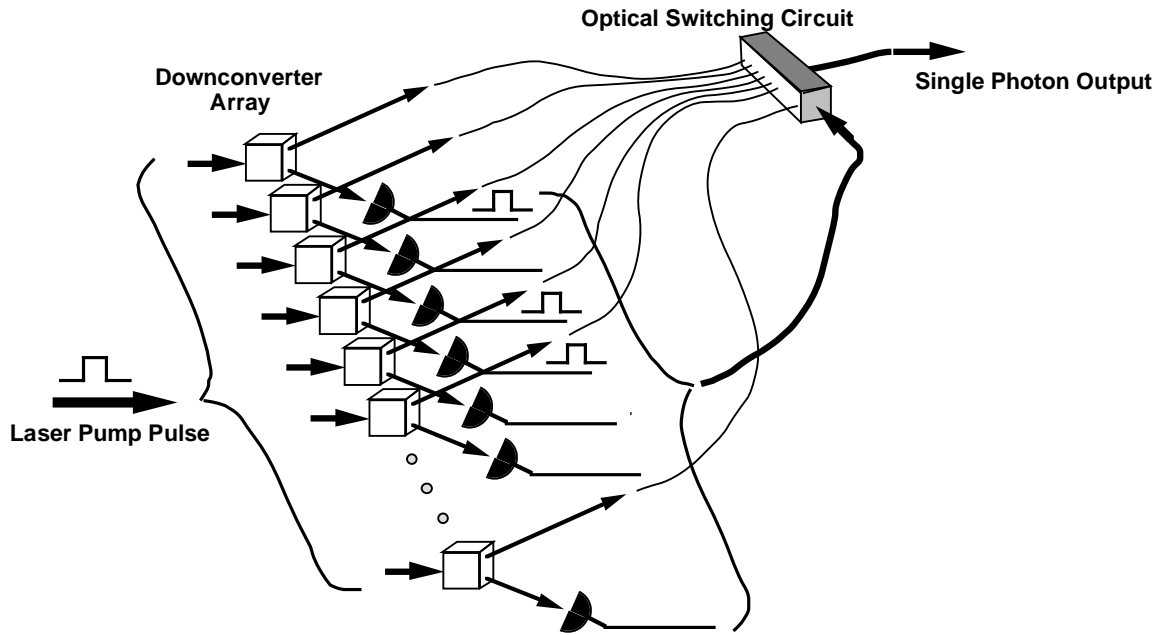}}%
  \caption{\label{fig1}
    General schematic for single photon source showing array of
    crystal downconverters each with the potential of producing a pair
    or pairs of photons. The each downconverter is shown with its own
    trigger detector. The information about which trigger detectors
    have fired is sent to the optical switching circuit to control
    which incoming lines are directed to the output line. Input line
    delays to allow the trigger information to arrive before the
    incoming photons are not shown.}
\end{figure}

\begin{figure}
  \centerline{\includegraphics{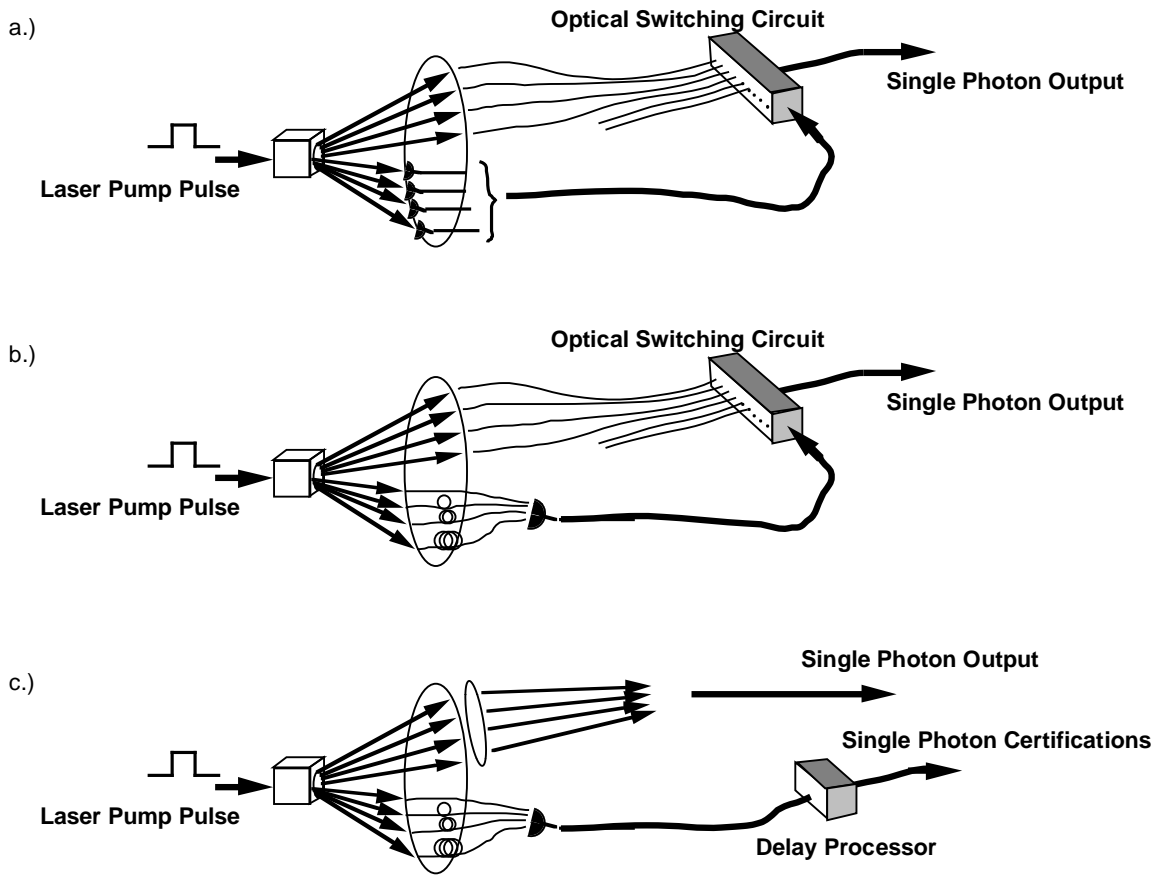}}%
  \caption{\label{fig2}
    a.) Single crystal implementation of scheme. b.) Single detector
    implementation of scheme. Sequential series of delay lines is seen
    leading to the trigger detector. c.) Implementation without the
    optical switching circuit. A lens is used to collect all the modes
    correlated to those seen by the trigger.}
\end{figure}

\section{Practical Implementation}

While the scheme just described is conceptually the simplest way
of presenting the method, there are a number of modifications that
can improve the efficiency, construction, and convenience of the
system. The first of these is that the array of downconverters can
be implemented with a single PDC crystal. This is possible because
while  phasematching requires a PDC photon pair to be constrained
to a plane containing the pump beam, the azimuthal angle of that
plane is not constrained. Thus, the PDC process produces light
distributed azimuthally around the pump laser direction (for type
I phasematching). So, each azimuthal plane can be thought of as a
separate downconverter. Thus the multiple PDC setup is achieved by
placing many detectors azimuthally in an annular pattern around
the pump direction of a single downconverter and collecting at the
correlated positions (Fig.~\ref{fig2}a).

A second modification allows the array of detectors to be replaced
by a single detector. By sending each of the potential herald
photons to an increasing sequence of delays and then directing the
delay outputs to a single detector, the timing of the detector
pulse indicates which of the input downconverters has created a
pair (Fig.~\ref{fig2}b). Of course, several of the downconverters
may produce a photon pair, but only the first photon herald
reaching the detector and causing it to fire is recorded. Detector
deadtime causes subsequent photons to be ignored. The timing of
the detector pulse is used to select which correlated photon
channel to direct to the output of system.

A third implementation  (Fig.~\ref{fig2}c) even eliminates the
output switching network circuit, while still maintaining a
significant advantage over both the conventional PDC and faint
laser photon sources. The output photons are simply collected with
a single lens output port. This most basic implementation allows
production of pulses with individual certainties of that pulse
containing exactly one photon. In other words, this source
provides single photons and a ``certificate" tied to each output
photon quantifying the likelihood that just one photon was
emitted. Some of these pulses can achieve a significantly higher
single photon certainty than is possible with the conventional
single photon source setups. This more complete characterization
of the emitted pulse and its tighter constraints on $P(>1)$, will
allow more efficient use of the emitted light. This reduces the
need for overhead tasks such as privacy amplification
\cite{ILM01}.

The basic reason that this arrangement can produce single photons
with lower probability of multiple photons is that the delay
system provides extra information about the photons produced. For
instance, in the cases where one of the longer delays happened to
cause the detector to fire, we know that all the prior delays did
not cause the detector to fire. If the detection quantum
efficiency is high,  it is very likely that there were no photons
in those modes coupled to those shorter paths. Thus $P(>1)$ is
greatly reduced because it is just the multiphoton probability for
only the last delay, rather than for all $N_D$ of them,  which has
a mean photon number of $\overline{n}$ vs. $\overline{n}/N_D$.

\section{Analysis}

We now quantify the advantage of this last and simplest scheme. We
will see this scheme results in production of photon pulses where
each pulse has its own individualized single photon certification,
and as expected, these certifications can be significantly better
than the uniform result obtained from the conventional
arrangement. To begin, we consider the standard PDC setup for
producing heralded single photons. The trigger detector registers
one photon of a pair and indicates the existence of the second
photon exiting the correlated channel. The collection optics for
that correlated channel are designed to collect, as close as can
be approximated, just the photons correlated to those seen by the
trigger detector. In this arrangement, both the trigger channel
and output channel are set up to collect only one mode of the
field \cite{KOW01}. When the trigger detector fires, one photon
has been received (assuming negligible dark counts), but we do not
know if additional photons were also present as the considered
detectors cannot distinguish a one-photon from a multi-photon
event. Given that the trigger detector has fired, the probability
that there were $n$ photons incident is:
  \begin{equation}
    \label{1}
    P^{\rm F}_{\overline{n},\eta}(n)= {{(1-(1-\eta)^{n}) \times
    P_{\overline{n}}(n)}\over {\sum_{i=0}^\infty (1-(1-\eta)^{i}) \times
    P_{\overline{n}}(n)}},
  \end{equation}
Note that $1-(1-\eta)^{n}$ is the probability of the detector
firing for $n$ photons incident and detector quantum efficiency,
$\eta$ defined as the probability of the detector firing when just
one photon is incident. We use Bose-Einstein statistics for the
probability,
$P_{\overline{n}}(n)={{\overline{n}^n}\over{(1+\overline{n})^{n+1}}}$,
of having $n$ photons emitted into a single mode of the PDC light,
given $\overline{n}$ \cite{ WAM95, GRT01}. The denominator of
Eq.~(\ref{1}) normalizes the distribution. Note that it does not
matter whether the summation includes $i$=$0$ or not, as the
probability of firing is zero if no photons are incident. We will
also need the probability  that there were one or more photons
incident if the detector did not fire. This function is given by
slight modifications of Eq.~(\ref{1}),
  \begin{equation}
    \label{2}
    P^{\overline{\rm F}}_{\overline{n},\eta}= {\sum_{k=1}^\infty
    (1-\eta)^{k} \times
    P_{\overline{n}}(k)\over {\sum_{k=0}^\infty (1-\eta)^{k} \times
    P_{\overline{n}}(k)}}\, ,
  \end{equation}
where $(1-\eta)^{k}$ is the probability of the detector not firing
if $k$ photons are incident and the numerator is the sum over all
the ways for the detector to not fire when photons are incident.
In the denominator, which is the total probability of not firing,
we now include $k$=$0$ , because that term represents a legitimate
way for the detector not fire. With this basis, we now describe a
system with a number of delay lines of increasing length placed
between the PDC crystal and the trigger detector. Each of these
delay line channels intercepts a single, but separate, mode of the
field. The output channel collection optic is also modified to
include the extra modes correlated to those of the additional
trigger modes.  Each of these trigger paths has a chance to cause
the trigger detector to fire, with the shortest path providing the
first chance, and the next longer path providing the next chance,
and so on. But once the trigger detector fires due to a photon in
a particular path, it cannot fire due to subsequent photons in the
longer delay paths. The timing of the trigger detector firing
relative to the pump pulse tells which delay path caused the
firing. Thus the result of a single pulse of the pump laser is
that either no trigger was produced or a trigger was produced and
we know which delay path produced it. This last piece of
information allows us to make a better determination of the
probability that the photon produced was a single photon.  If the
photon that causes the trigger to fire is one of the later delay
paths, we will have a much lower likelihood of there being more
than one photon. We now calculate this likelihood as a function of
which delay path caused the firing.

Suppose we have a system of $N_{\rm D}$  delay paths and we have
determined that the $i^{\rm th}$ delay path caused the firing. We
then obviously know that all the previous delay paths did not
cause the detector to fire, so the probability that only one
photon was incident into the entire system of delay paths is given
by
  \begin{equation}
    \label{3}
    P_{\overline{n},\eta,N_{\rm D}}(i)={\left({1-P^{\overline{\rm
    F}}_{{\overline{n}\over{N_{\rm D}}},\eta}}\right)^{i-1}}
    P^{\rm F}_{{\overline{n}\over{N_{\rm D}}},\eta}(1)
    \left({P_{\overline{n}\over{N_{\rm D}}}(0)}\right)^{N_{\rm D}-i}
    ,
  \end{equation}
where $P^{\overline{\rm F}}_{{\overline{n}\over{N_{\rm D}}},\eta}$
is the probability that photons were incident if the detector did
not fire. The first factor is the probability that all delays
prior to the $i^{\rm th}$ delay did not fire due there being no
photons in those paths. The second factor is the probability that
the $i^{\rm th}$ delay that caused the firing contained exactly
one photon. The last term is the probability that all the
subsequent delay paths contained no photons. Note that
$\overline{n}$ is the mean photon rate for the overall system, so
the rate for a single mode is  $\overline{n}\over{N_{\rm D}}$.
Then we use Eq.~(\ref{1}) with $n$=1 for a Bose-Einstein
distribution in each of the modes collected of Fig.~\ref{2}c.

We need to calculate not only the probability of a single photon
when delay $i$ causes the trigger to fire, but also the likelihood
of that event occurring. This calculation is needed to verify that
while a particular result may give a high probability of being a
single photon, the probability of having that result is not
impractically small. This probability of a particular delay
causing the trigger to fire is given by
  \begin{equation}
    \label{4}
    P_{i}=\left(\sum_{n=0}^\infty (1-\eta)^{n} P_{\overline{n}
    \over{N_{\rm D}}}(n)\right)^{i-1}
    \times \left(1-\sum_{n=0}^\infty (1-\eta)^{n} P_{\overline{n}
    \over{N_{\rm D}}}(n)\right)\, ,
  \end{equation}
Here the first factor is the probability of that all the paths
prior to  $i^{\rm th}$ delay did not fire and the second factor is
the probability that the  $i^{\rm th}$ delay did fire (written as
1 minus the probability that the  $i^{\rm th}$ delay did not
fire). Also, we must include the possibility that no photon was
detected via any delay path, i.e. that the trigger did not fire at
all. The likelihood of this occurring is given by
\begin{equation}
  \label{5}
  P_{0}=\left(\sum_{n=0}^\infty (1-\eta)^{n} P_{\overline{n}
  \over{N_{\rm D}}}(n)\right)^{N_{\rm D}}\, ,
\end{equation}
We can now substitute the Bose-Einstein distribution $
P_{\overline{n} \over{N_{\rm D}}}(n) $ into the above equations
and simplify. Eq.~(\ref{3}) becomes
  \begin{equation}
    \label{6}
    P_{\overline{n},\eta,N_{\rm D}}(i)=\left(N_{\rm D}\over{\overline{n}+N_{\rm D}}
    \right)^{i-1+N_{\rm D}} \times
    \left({N_{\rm D} +\eta \overline{n}}\over{\overline{n}+N_{\rm D}}
    \right)^{i} ,
  \end{equation}
and the probability of a particular delay causing the trigger to fire becomes
  \begin{equation}
    \label{7}
    P_{i}={{\eta \overline{n}}\over{N_{\rm D}}} \left({N_{\rm D}
    }\over{\eta \overline{n}+N_{\rm D}}
    \right)^{i}\, ,
  \end{equation}
while
  \begin{equation}
    P_{0}=\left({N_{\rm D}}\over{\eta \overline{n}+N_{\rm D}}
    \right)^{N_{\rm D}}
    \label{8}.
  \end{equation}

Figure (\ref{3}a) shows the functional form of these probabilities
(of Eq.~(\ref{6})), where each line of the fan shaped family of
curves represents a system of $N_{\rm D}$ delays. (Figure \ref{3}b
showing the probability of a single delay event occuring is
discussed later.) Each point on a given line corresponds to the
trigger firing at a particular $i^{\rm th}$ delay out of a set of
$N_{\rm D}$. The point's value is the ``single photon
certification'' - the probability that this event indicates that
exactly one photon pair exists in the system (one photon in the
trigger channel and one in the output channel). It can be seen
that for $N_{\rm D}$ of more than a few, and with high $\eta$, we
can have emission events with single photon probabilities around
90\%, greatly exceeding the conventional arrangement results for
the same $\eta$ and $\overline{n}$. The conventional result is
represented by the single  $N_{\rm D}$ =1 point (i.e. the standard
PDC setup with one trigger path).
  \begin{figure}[ht]
    \centerline{\includegraphics{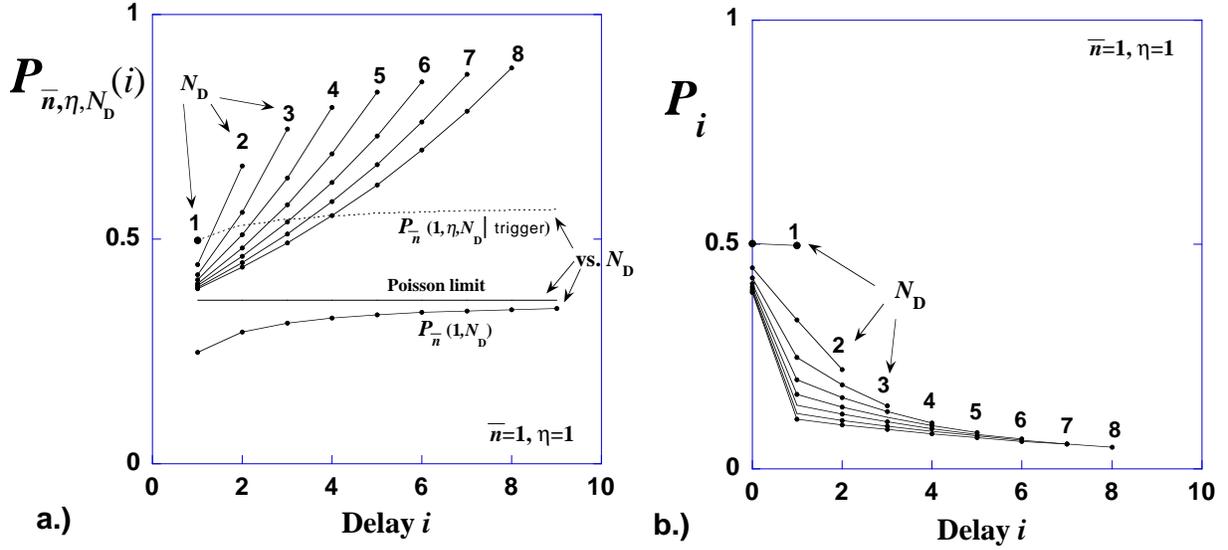}}%
    \caption{\label{fig3} a) Plot of single photon ``certifications,''
    $P_{\overline{n},\eta,N_{\rm D}}(i)$ as a function of the
    $i^{\rm th}$ delay firing. The fan of curves labeled 1-8 are
    the probabilities of exactly one photon being produced given that
    the $i^{\rm th}$ delay, in a system of $N_{\rm D}$, caused the
    trigger to fire for $\eta$,$\overline{n}$=1. The lowest curve is
    the total probability for a system of $N_{\rm D}$ delays to
    produce a single photon per pump pulse with the Poisson limit
    shown just above. The dashed curve above the Poisson limit is the
    probability that the emitted light is a single photon given that
    the trigger did fire. (For these last 3 curves, the $x$-axis is
    $N_{\rm D}$  rather than triggered delay.)  b)  Probability of
    that the $i^{\rm th}$ delay will cause the trigger to fire.}
  \end{figure}

We have calculated two additional probabilities: the overall
probability of producing a single photon (Eq.~(\ref{9})), and that
same probability given that the trigger did fire (Eq.~(\ref{10})).
The first is obtained by taking the product of the probability of
a particular delay event being due to a single photon and the
probability of that event occurring, and then summing over all
possible types of events ($i$=1,2,... $N_{\rm D}$). Added to this
is the probability for the case where no delays fire (so no
heralding is present), but there is still a single photon
(undetected) in one delay line:
  \begin{equation}
    P_{\overline{n}}(1,\eta,N_{\rm D})=\sum_{i=1}^{N_{\rm D}}
    P_{\overline{n},\eta,N_{\rm D}}(i) \times P_{i} +
    N_{\rm D}  (P_{{\overline{n}\over{N_{\rm D}}}}(0))^{N_{\rm D}-1}
    (1-\eta)  P_{{\overline{n}\over{N_{\rm D}}}}(1)
    \label{9}.
  \end{equation}
The second probability is obtained by eliminating the case where
the trigger did $not$ fire and renormalizing:
  \begin{equation}
    P_{\overline{n}}(1,\eta,N_{\rm D}|trigger)={{
    \sum_{i=1}^{N_{\rm D}} P_{\overline{n},\eta,N_{\rm D}}(i) \times P_{i} }
    \over{1-P_{0}}}
    \label{10}.
  \end{equation}
Via some algebra we obtain:
  \begin{equation}
    P_{\overline{n}}(1,N_{\rm D})=
    \overline{n}  \times
    {\left({N_{\rm D}\over{\overline{n}+N_{\rm D}}}\right)^{1+N_{\rm D}}}
    \label{11}
  \end{equation}
and
  \begin{equation}
    {P_{\overline{n}}(1,\eta,N_{\rm D}|trigger)}=
    {{{\overline{n}
    \eta\left({N_{\rm D}\over{\overline{n}+N_{\rm D}}}\right)^
    {1+N_{\rm D}}}}\over{1- {\left({N_{\rm D}\over{\overline{n} \eta
    +N_{\rm D}}}\right)^{N_{\rm D}}}
    }}
    \label{12}
  \end{equation}
for a Bose-Einstein distribution in each of the modes collected in
Fig. 2c. These two results are also shown on Fig 3a for $\eta$=1.
(Note, $P$ is independent of $\eta$ in Eq.~(\ref{11}), a
reflection of the fact that it does not involve the herald, or
trigger, detector.)  For these curves, the abscissa is $N_{\rm D}$
rather than delay $i$.

It is also instructive to calculate the probabilities in Eqs.~(9)
and (10) assuming the heralded photons follow a Poisson, rather
than Bose-Einstein, distribution in each mode. Then Eq.~(\ref{9})
simply returns the Poisson probability of exactly one photon (the
Poisson limit):
  \begin{equation}
    P_{\overline{n}}(1)=\overline{n} ~ e^{-\overline{n}}
    \label{13},
  \end{equation}
while Eq.~(\ref{10}) becomes:
  \begin{equation}
    {P_{\overline{n}}(1,\eta|trigger)}=
    {{{{\overline{n}
    \eta e^{-\overline{n}}}}}\over{1- e^{-\overline{n}\eta}}} .
    \label{14}
  \end{equation}
These are independent of $N_{\rm D}$, as one would expect, because
a collection of Poisson subsystems taken together yields a result
with Poisson statistics for the entire system.

  \begin{figure}[ph]
    \centerline{\includegraphics[totalheight=7in]{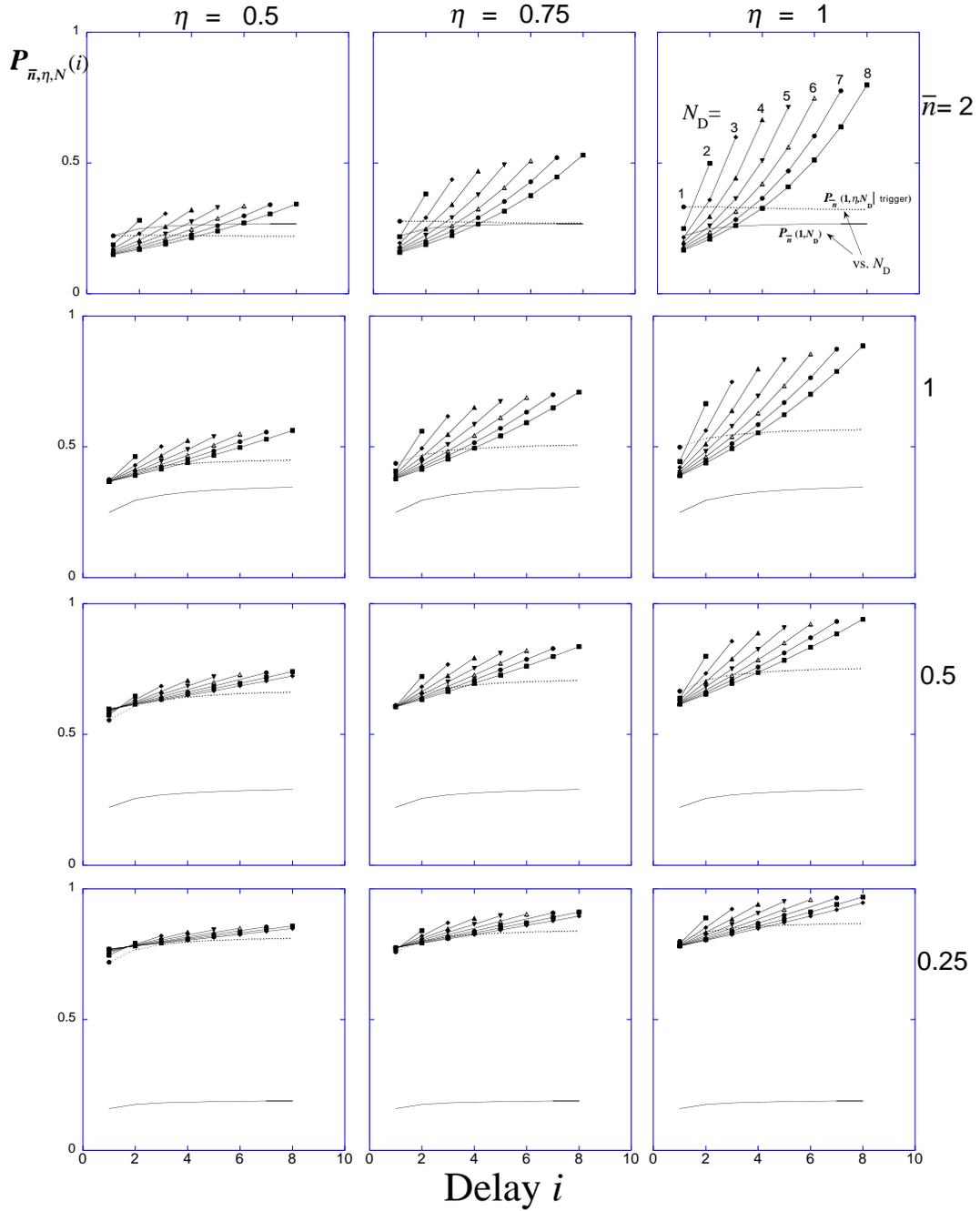}}
    \caption{\label{fig4} Matrix of graphs single photon probabilities
    similar to fig 3a for a range of values of $\overline{n}$  and $\eta$.
    The columns correspond to $\eta$'s of 0.5, 0.75, and 1 from left
    to right, while the rows correspond to  $\overline{n}$'s of 2, 1,
    0.5, 0.25 from top to bottom.}
  \end{figure}

For large $N_{\rm D}$, the probabilities of Eqs.~(\ref{11}) and
(\ref{12}), calculated for a thermal source, approach the results
that would be obtained for Poisson-distributed photons as given by
Eqs.~(\ref{13}) and (\ref{14}). In particular,  the unheralded
single-photon probability $P_{\overline{n}}(1,N_{\rm D})$
asymptotically approaches the Poisson limit of Eq.~(\ref{13}) for
$N_{\rm D}\rightarrow \infty$, as seen in Fig. 3a. This tending of
a collection of many single-mode Bose-Einstein subsystems toward
the Poisson result is also expected, in that the more independent
subsystems are included in the sum, the more the individual events
in the system are independent of each other, which is the
definition of Poisson statistics. The higher values of
$P_{\overline{n} }(1,\eta,N_{\rm D}|trigger)$ seen in Fig. 3a
indicate the advantage of heralded Bose-Einstein-distributed
photons over those from a faint laser, whose single-photon
probability is given by the Poisson limit. An even bigger
advantage would occur if it were possible to have heralded,
Poisson-distributed photons: the above analysis yields
qualitatively the same shaped curves for the single-photon
certifications $P_{\overline{n},\eta,N_{\rm D}}(i)$ as in Fig.~3a,
but with even higher probability values in that case.

As previously mentioned, we must also verify that the events with
high single photon probabilities or ``certifications" have
reasonable likelihoods of occurring. Fig. \ref{fig3}b shows the
probabilities of occurrence for each type of event shown in Fig.
\ref{fig3}a. (For comparison we have included the probability of
the trigger detector not firing due to any of the delay channels.
These are the points plotted at delay=0.) While the likelihood of
the later delay events occurring is lower than the earlier delay
events, the dependence is not particularly strong. For instance in
the $N_{\rm D}$=8, $\overline{n}$=1 case the falloff from delay-1
events to delay-8 events is only a factor of about 2.

Figure \ref{fig4} is a matrix of graphs similar to Fig. \ref{3}a
for a range of values of $\overline{n}$  and $\eta$. Looking from
left to right, we see that as  $\eta$  approaches unity, we
achieve the best single photon certifications. This is because a
high  $\eta$ means that the system will provide more complete
information about what has happened. For example with a high
$\eta$, an instance of the trigger not firing means with high
certainty that no photon was incident, while a low   $\eta$
decreases our certainty of that event. Highlighting the advantage
of this method, fig. 4 also shows that in almost all cases, the
multiplexed heralded system presented in this paper significantly
surpasses the single photon probability of a faint laser as
described by the Poisson limit. For each value of $\overline{n}$
this limit is indicated on the figure by the level of the
asymptote to the $P_{\overline{n}}(1,N_{\rm D})$ curve.

Figure \ref{fig4} also shows that increasing $\overline{n}$,
increases the spread of the certifications, while decreasing the
maximum single photon certification possible. So we can see that
there is a trade off between having high certainty single photons
and overall single photon rate. In fact,
$P_{\overline{n}}(1,N_{\rm D})$ does not continue increasing with
increasing $\overline{n}$. This is clearly seen in Fig. \ref{5},
which shows $P_{\overline{n}}(1,N_{\rm D})$ for systems with
varying $N_{\rm D}$, as a function of the mean overall rate of
photon generation per pulse. The maximum occurs for all systems at
$\overline{n}$=1. This is the balancing point between a low photon
rate that reduces the number of single photon events and a high
photon rate that increases the number of multiple photon events.
This tells us the best rate to operate the system to maximize
single photon events, although it will not necessarily provide the
highest single photon certainties. But we can still achieve higher
single photon certifications than is possible conventionally,
while maximizing single photon rates.

  \begin{figure}
    \centerline{\includegraphics{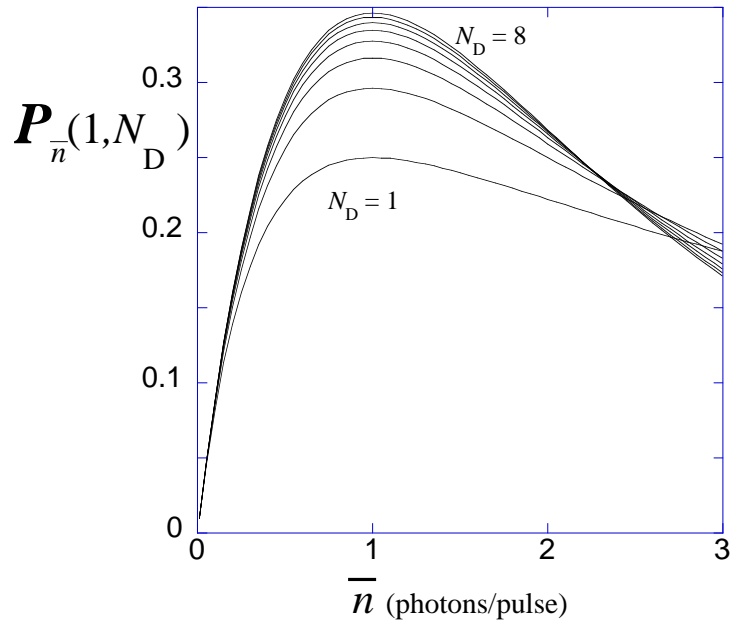}}
    \caption{\label{fig5} Total probability of producing a single
    photon vs. mean overall rate of photon generation per pulse for
    systems with $N_{\rm D}$ from 1 to 8 with $\eta$=1.}
  \end{figure}

For comparision of the pulsed ``single photon'' systems using a
faint laser, conventional PDC source, and our multiplexed PDC
source, we calculate the overall single photon production
probability. For a faint laser, the maximum possible fraction of
single photon events is 37\%. For a conventional PDC source, the
single-photon fraction is limited to only 25\% because the
Bose-Einstein statistics encourage bunched photons; however
because the PDC source provides a herald, the no-photon emission
event can be eliminated bringing the single photon fraction up to
50\%. Then, using the full multiplexed PDC setup presented here,
fractions much closer to 100\% can be achieved, while even the
simplest multiplexed version of Fig. 2c allows a single photon
fraction of 57\% (for the case $N_{\rm D}=8$, $\overline{n}$=1,
$\eta$=1).

\section{Conclusion}

We have shown a way to decouple the probabilities of producing a
single photon and the probability of producing more than one
photon, using an array of parametric downconverters.  By doing so,
we can construct a better approximation of a true single photon
on-demand source than is possible using a conventional single PDC
setup or a faint laser source \cite{BLM00, GRT01}. In principle
this method could achieve an arbitrarily close approximation of a
single photon on demand source. We have also analyzed a version
that, while greatly simplifying the construction of an actual
device, retains a significant amount of the benefit of the
original concept. The setup would produce single photons with
individual certifications that the photons produced are actually
the desired single photons. Such a better-defined single photon
source will allow for better use of quantum channel resources in a
cryptographic system by reducing the need for overhead tasks such
as privacy amplification, as well as having implications for the
field of quantum computation. As photon counting becomes more
convenient at telecom wavelengths, we expect that integrated all
solid state implementations of these schemes will be made even
easier, and we will have truly achieved the dream of a convenient
source of single photons on demand. We are currently working on
experimental implementations.



\end{document}